\documentclass[eprint,amsmath,amssymb,aps, superscriptaddress]{revtex4-2}

\usepackage{graphicx}
\usepackage{graphics}
\usepackage{dcolumn}
\usepackage{bm}
\usepackage{amsmath}
\usepackage{amssymb}
\usepackage{float}

\usepackage{verbatim}
\usepackage{booktabs}
\usepackage{xcolor}
\usepackage{url}
\usepackage[pdfencoding=auto]{hyperref}

\newcommand{\eqn}[1]{\begin{align}#1\end{align}}
\newcommand{\bs}[1]{\boldsymbol{#1}}

\newcommand{\pare}[1]{\left( #1 \right) }

\newcommand{\fr}[2]{\frac{#1}{#2}}

\newcommand{\mc}[1]{\mathcal{#1}}

\definecolor{darkgreen}{rgb}{0, 0.5, 0.05}

\newcommand{\rev}[1]{{\color{black}#1}}

\newcommand{\deleted}[1]{}

\def\dd{\mathrm{d}}

\def\bx{\bs{x}}

\def\be{\bs{e}}
\def\bu{\bs{u}}

\def\bmu{\bs{\mu}}

\def\bG{\bs{G}}

\begin{document}


\title{Viscosity ratio across interfaces controls the stability and self-assembly of microrollers}

\author{Blaise Delmotte}
\email{blaise.delmotte@cnrs.fr}
 \affiliation{LadHyX, CNRS, Ecole Polytechnique, Institut Polytechnique de Paris, 91120 Palaiseau, France}

\begin{abstract}
We investigate the individual and collective dynamics of torque-driven particles, called microrollers, near fluid-fluid interfaces. We find that the viscosity ratio across the interface controls the speed and direction of the particles, their relative motion, the growth of a fingering instability and the self-assembled motile structures that emerge from it. By combining theory and large scale numerical simulations, we show how the viscosity ratio across the interface governs the long-range hydrodynamic interactions between particles and thus their collective behavior.
\end{abstract}

\pacs{Valid PACS appear here}
                    
\maketitle
\begin{figure}[t]
    \centering
    \includegraphics[width=1\columnwidth]{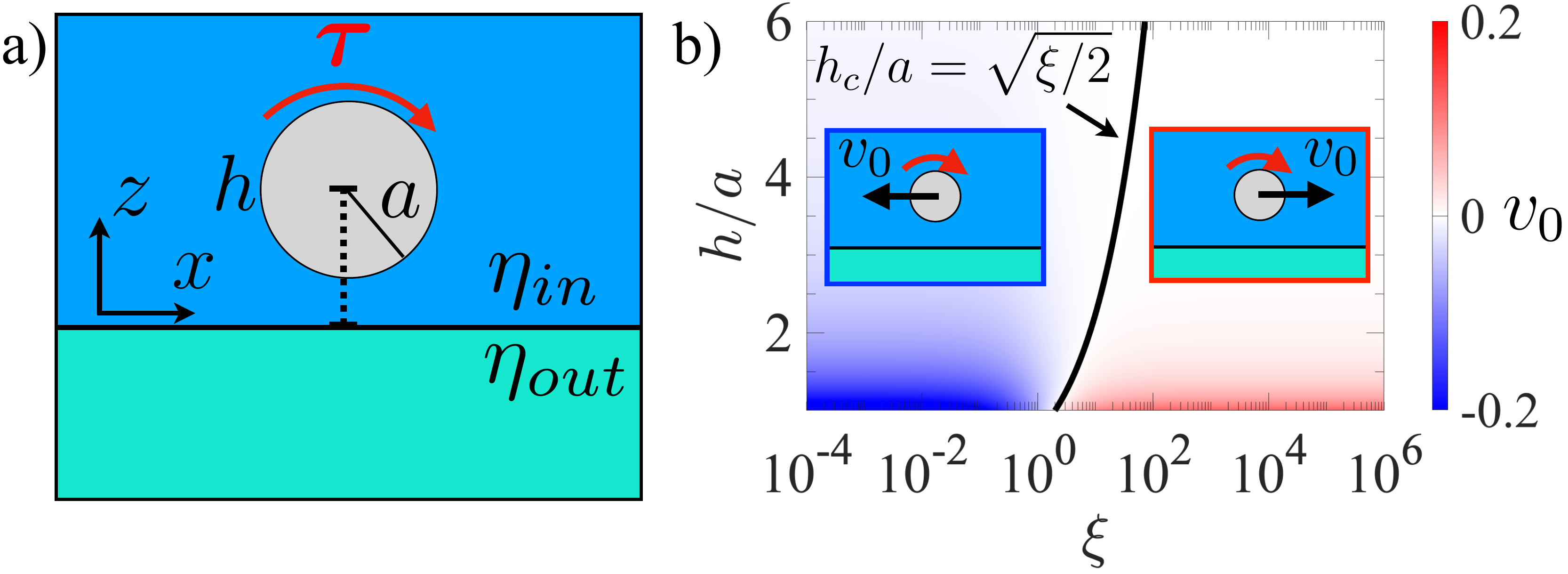}
    \caption{a) Schematic of a torque-driven particle above a fluid-fluid interface. b) Self-induced velocity of the particle $v_0$ in response to a constant torque $\tau$ as a function of $\xi=\eta_{out}/\eta_{in}$ and normalized height $h/a$.} 
    \label{fig:fig1}
\end{figure}
\begin{figure}[t]
    \centering
    \includegraphics[width=0.7\linewidth]{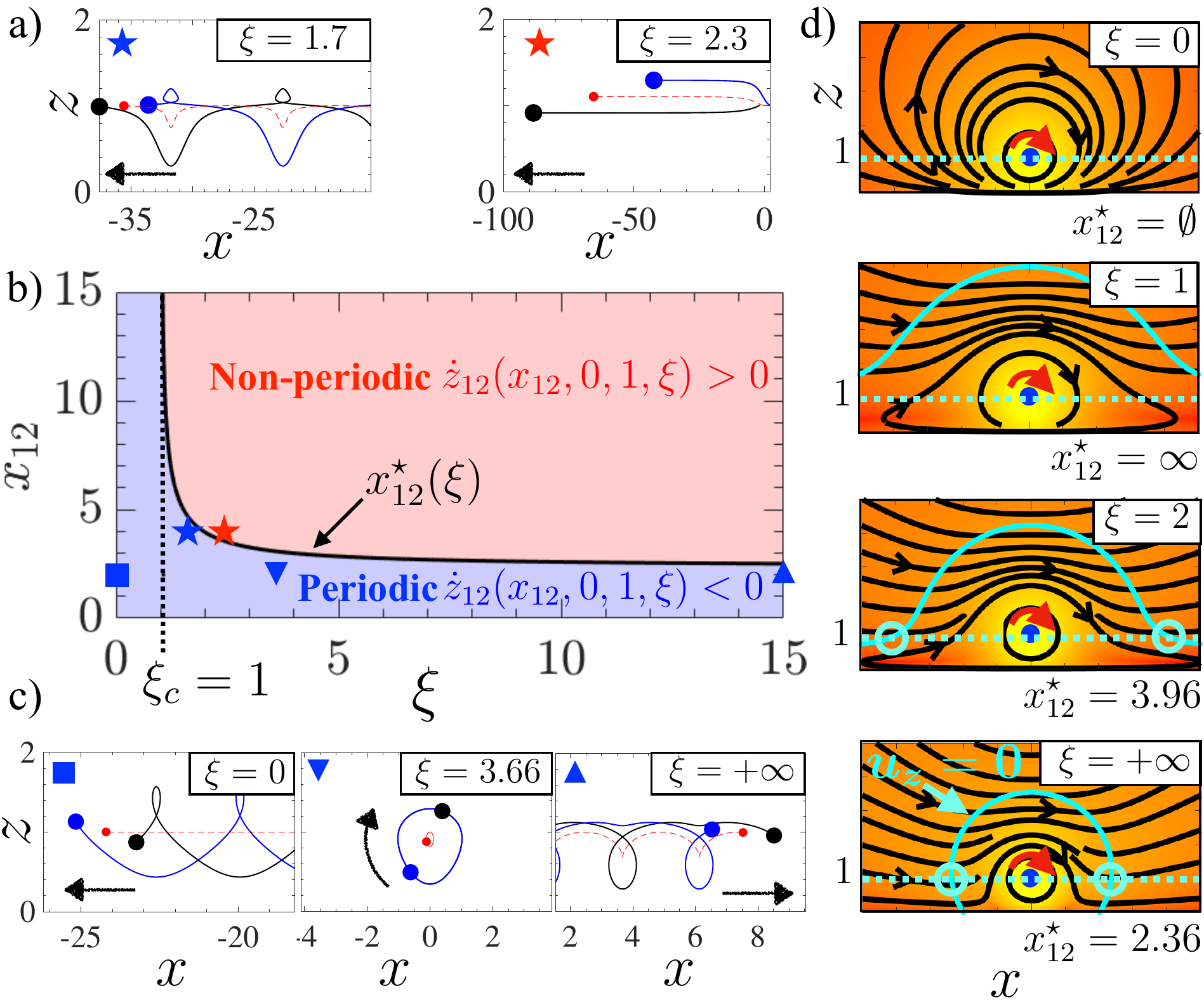}    
    \caption{a) Effect of $\xi$ on the periodicity of trajectories of two rotlets (blue and black disks) with identical initial separation: $x_{12}^0=4$. Red disk and dashed line: position and trajectory of the center of mass ($x_C,z_C$). Black arrow:  direction of motion of $x_C$. b) Phase diagram of rotlet orbits near the critical points as a function of $\xi$ and $x_{12}$. Black line: critical distance $x_{12}^{\star}$. c) Effect of $\xi$ on the direction of motion of a pair with $x_{12}^0=2$. d) Streamlines and flow field around a rotlet in the $(x,z)$-plane for various $\xi$. Cyan solid line: iso-value $u_z=0$. Cyan dotted line: $z=1$. Cyan circles: position of the critical horizontal separation $x_{12}^\star$, reported below each panel.
    }
    \label{fig:fig2}
\end{figure}

The motion of small driven particles near interfaces is observed in numerous biological and artificial systems \cite{Lauga2016,Bechinger2016,Fei2017}. Examples include droplets filled 
 with self-propelled particles, where hydrodynamic couplings between the particles and the surrounding fluid-fluid interface induce collective motion and translation of the droplet \cite{Gao2017,Vincenti2019,Singh2020,Huang2020,Rajabi2021,Kokot2022}. Navigation near interfaces also happens in natural environments with air-water interfaces, such as marine foams \cite{Roveillo2020} or films \cite{DiLeonardo2011,Morse2013,Mathijssen2016b}, and in biomedical environments where microswimmers move near soft vessel walls, visco-elastic media (biofilms, gels, tissues) or mucus layers \cite{Suarez2006,Lopez2014,Alapan2020}. \\
Among them, spinning particles, such as torque-driven colloids (also called microrollers), use rotation-translation couplings near surfaces to self-propel. Synthetic microrollers are actuated by an external magnetic field rotating about an axis parallel to the interface. 
The orientation of the magnetic field can be varied over time to guide these particles in a variety of environments. Thanks to their steerability and to the strong long-range flows they generate, they offer promising perspectives for particle micromanipulation, fluid pumping and drug delivery in microfluidic and biological systems \cite{Kavvcivc2009,Martinez2015PRA,Martinez2018, Driscoll2018, Chamolly2020,Qi2021,Bozuyuk2022,Demirors2021}. Above a rigid wall, hydrodynamic interactions between microrollers induce a variety of collective motions such as periodic leapfrog orbits \cite{Martinez2018b,Delmotte2019}, the formation of fast moving layers \cite{Martinez2015,Sprinkle2020,Junot2021}, the emergence of dense fronts \cite{Delmotte2017} and the growth of a fingering instability that releases stable motile clusters \cite{Delmotte2017b,Driscoll2017}. However, despite the rich and well documented dynamics observed above rigid boundaries, little is known about their collective behavior near interfaces.\\
In this Letter, we combine theory and  simulations to investigate the dynamics of microrollers  near fluid-fluid interfaces. We show how the viscosity ratio across the interface, denoted $\xi$, modifies the flow around the spinning particles, with amazing  consequences on their individual and collective motion.
At the individual level, we find that $\xi$ controls
the direction and speed of a single microroller. At the
pair level, $\xi$ acts as an order parameter on their relative motion that determines the existence of periodic orbits. At the collective level, $\xi$ controls both the growth rate of the fingering instability initially observed above a no-slip surface, and the self-assembly process that leads to the emergence of motile clusters.
As discussed in the conclusions, the ability to control the macroscopic response of these active suspensions opens promising perspectives for microfluidics applications.\\
We consider the motion of small torque-driven spherical particles with radius $a$ suspended in a fluid with dynamic viscosity $\eta_{in}$ above a fluid-fluid interface (Fig.\ \ref{fig:fig1}a).  The outer fluid on the other side of the interface has dynamic viscosity $\eta_{out}$ which can vary from zero (e.g.\ air) to $+\infty$ (a rigid wall). We denote $\xi = \eta_{out}/\eta_{in}$  the viscosity ratio between the two fluids across the interface. In this work we consider particles with large contact angles that are not adsorbed to the surface (such as  paramagnetic beads above air/water surfaces 
\cite{Martinez2015}). In typical experiments the particles are micron-sized ($a=O(1)\mu$m) and suspended in water ($\eta_{in}=10^{-3}$Pa$\cdot$s). They rotate in synchrony with an external magnetic field with frequency $f = O(10)$Hz thanks to a magnetic torque aligned with the $y-$axis denoted $\tau$.  The corresponding capillary number is  \rev{Ca$=\eta_{in}u/\gamma \approx 10^{-6}-10^{-5}$}, where $u=\omega a=2\pi f a$ is the maximal fluid velocity, reached on the particle surface due to the no-slip condition, and \rev{$\gamma = O(10^{-2})$N/m} is the typical surface tension between the two phases. Owing to the very small value of Ca and to the small size of the particles, the interface can be \rev{approximated} as flat and nondeformable \rev{for all values of $\xi$ \cite{Lee1979}}. In this limit, one can compute the flow field around the particles using  the Green's function $\bG$ of the Stokes equations which, by linearity, is given by~\cite{Swan2007}: 
\eqn{
\bG =  1/(\xi+1)\bG^{FS} + \xi/(\xi+1)\bG^W
\label{eq:Green}
}
where  $\bG^{FS}$ is the Green's function of a domain bounded by a flat free-surface  and $\bG^W$ a domain bounded by a flat no-slip wall, both of which have known  analytical expressions based on image systems~\cite{Blake1974,Lee1979}\footnote{\rev{Note that if the interface is not flat, e.g.\ near a small droplet, \eqref{eq:Green} involves additional terms that cannot be written as simple linear combinations of $\bG^{FS}$ and $\bG^W$ \cite{Fuentes1989}.}}.
Changing $\xi$ from 0 to $+\infty$ in \eqref{eq:Green} transitions smoothly from a free-surface to a no-slip wall.\\
\textit{\textbf{Single particle motion.}--}
Using the Faxen formulae for the motion of a spherical particle in a non-uniform flow \cite{Kim1991}, the far-field approximation of the self-induced velocity of a single microroller above the interface, is given by \cite{SI}
\eqn{
v_0(\xi,h/a) =  \frac{1}{4} \pare{\frac{a}{h}}^2\left[-\frac{1}{\xi +1} + \frac{\xi}{\xi+1}\frac{1}{2}\pare{\frac{a}{h}}^2\right],
\label{eq:v0}
}
where the velocity has been normalized by $\tau/8\pi\eta a^2$.
Interestingly, the sign of the velocity in \eqref{eq:v0} depends both on the viscosity ratio \textit{and} the particle height (Fig.\ \ref{fig:fig1}b). Below a critical value $\xi_c=2$ the velocity is always negative, even though the applied torque is clockwise, and decays as $(a/h)^2$ for $\xi=0$. This surprising backward motion has already been observed experimentally \cite{Martinez2015} and is due to the fact there is no velocity gradient on the free-surface underneath. The particle therefore experiences more viscous stress on its upper side, where velocity gradients are larger, than on the lower side. As a result, the particle ``rolls" on the liquid above, which resists more against rotational motion, and thus moves backwards. When $\xi\geq\xi_c$, there is a critical height $h_c=\sqrt{\xi/2}a$ at which viscous stresses balance between both sides such that the particle rotates in place, i.e.\ $v_0(\xi>\xi_c,h_c) = 0$. $v_0$ is positive below $h_c$ and negative above.  In the limit of a no-slip wall, $\xi\rightarrow+\infty$, $v_0$ is always positive and decays as $(a/h)^4$. These results show that both the direction and speed of a rotating particle can be controlled with the viscosity ratio.\\
\textit{\textbf{Pair dynamics.}--}
When adding another particle in the system, the trajectories exhibit a richer dynamics than the individual motion described above:  particles can follow a variety of periodic orbits, change direction or end up translating at a steady speed (Fig.\ \ref{fig:fig2}a-c). We consider two torque-driven particles, lying in the  $(x,z)$-plane vertical to the floor and separated by a distance $r_{12} = \sqrt{x_{12}^2 + z_{12}^2}$, where $x_{12}=x_1-x_2$ and $z_{12}=z_1-z_2$. For simplicity, we focus on the limit $a\ll \min(r_{12},z_1,z_2)$, where the particles are considered as point-torques (rotlets) so that the equations of motion can be written compactly as a simple dynamical system with three degrees of freedom \cite{SI}:  
\eqn{
\left[\begin{array}{c}
\dot{x}_{12}\\
\dot{z}_{12}\\
\dot{z}_C
\end{array}\right] = 
\left[\begin{array}{c} 2z_{12}\left(\frac{1}{r_{12}^3} - \fr{\xi}{\xi+1} \frac{2x_{12}^2-4z_C^2}{R_{12}^5}\right) + \delta v_0\\
 2x_{12}\left(\frac{1}{R_{12}^3} - \frac{1}{r_{12}^3} +  \fr{\xi}{\xi+1}\frac{12 z_C^2}{R_{12}^5}\right) \\
\fr{\xi}{\xi+1} \frac{6 x_{12}z_{12}z_C}{R_{12}^5}
 \end{array}\right]
 \label{eq:dyn_2particles}
}
where $z_C = (z_1+z_2)/2$, and $R_{12} = \sqrt{x_{12}^2 + 4z_C^2}$. $\delta v_0 = -\frac{1}{4}\fr{1}{\xi+1}\pare{(z_C+z_{12}/2)^{-2}-(z_C-z_{12}/2)^{-2}}$ is the difference between the self-induced velocities in the point-particle limit. Here lengths have been rescaled by the initial height of the system $l_c=z_C^0 = z_C(t=0)$ and time by $t_c=8\pi\eta l_c^3/\tau$.
The dynamical system \eqref{eq:dyn_2particles} has two critical saddle  points with coordinates  $\bx^{\star}_{\pm} = ( x_{12}=\pm x_{12}^{\star},z_{12}=0,z_C=1)$, where $x_{12}^{\star}\neq 0$ is a zero of  $\dot{z}_{12}(x_{12},0,1,\xi)$ \cite{SI}.
  \begin{figure}[t]
    \centering
    \includegraphics[width=1\linewidth]{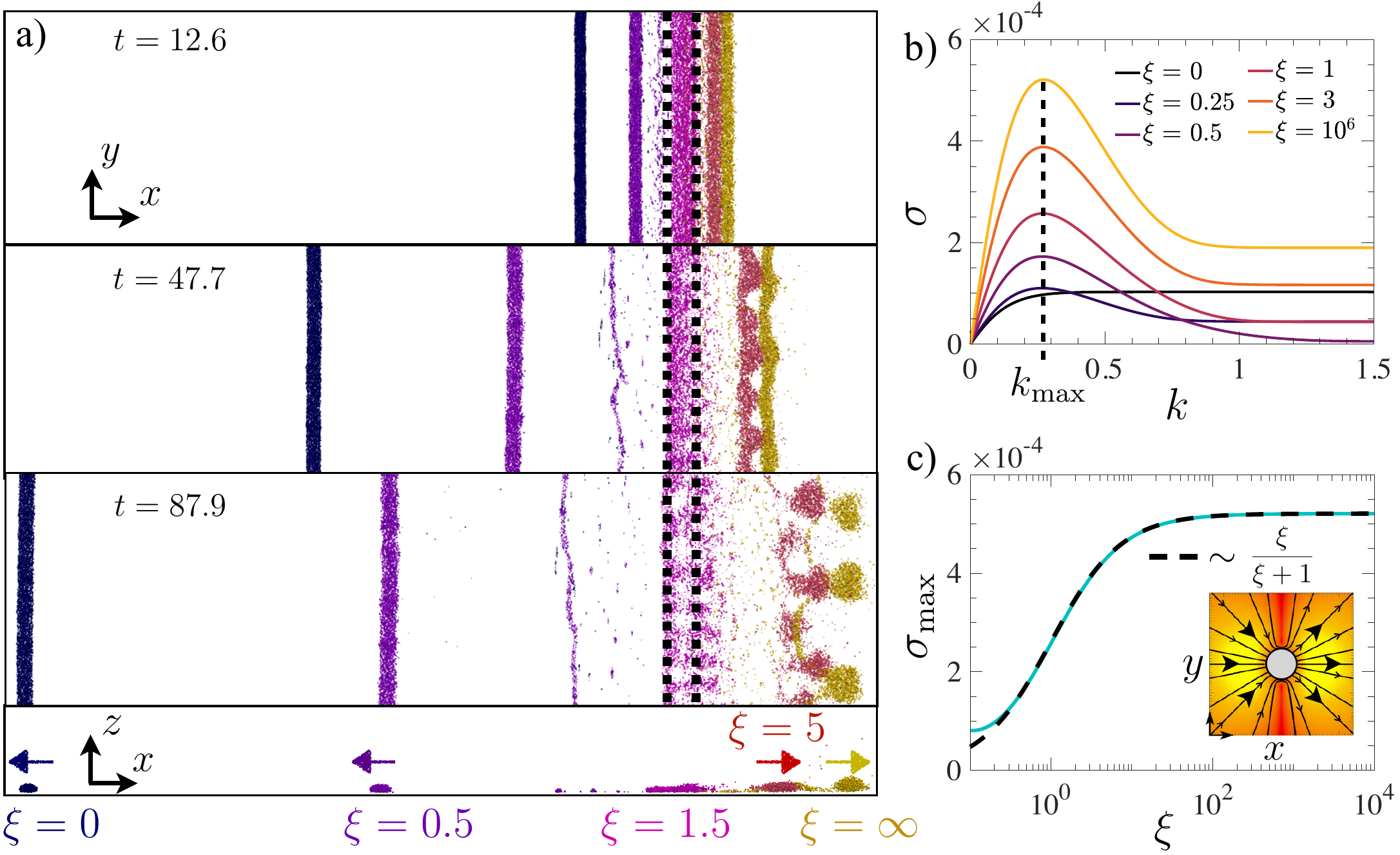}    
     \caption{a) Time-evolution of 10,000 microrollers initially uniformly distributed in a monolayer (delimited by the black dotted lines)  above the interface. Each color corresponds to a different value of $\xi \in \{0, 0.5, 1.5, 5, +\infty\}$ simulated independently. Bottom panel: side view at $t=87.9$. Arrows: direction of motion. b) Growth rate of the two line model \eqref{eq:conserv_lines}-\eqref{eq:conv_prod1} for various $\xi$. Dashed line: fastest growing mode $k_{\max}$. c) Maximum growth rate (solid cyan line) as a function of $\xi$. Dashed black line:  prefactor $\sim \xi/(\xi+1)$ due to the transverse flows generated above a no-slip boundary (inset).
     } 
    \label{fig:fig3}
\end{figure}
These critical points separate periodic (leapfrog) and non-periodic orbits, as shown on Fig.\ \ref{fig:fig2}b.  
 Interestingly, their position, $x_{12}^{\star}$, is controlled by the viscosity ratio $\xi$.  To visualize this dependency, we examine the changes in the topology of the flow field around a spinning particle as a function of $\xi$ (Fig.\ \ref{fig:fig2}d). When $\xi=0$ all the streamlines in the domain are closed and circular. As $\xi$ increases, the region of closed streamlines contracts in the vicinity of the particle, while open streamlines, which push the fluid forward, occupy more space. To understand the effect of the flow on the periodicity of the orbits we place the particles at the critical point $x_{12}=x_{12}^{\star}$ where  $z_1= z_2 = 1$ and we assume $x_1>x_2$. At this position their relative, and thus absolute, vertical velocity vanishes  $\dot{z}_{12}=0\Rightarrow \dot{z}_1= \dot{z}_2 = \dot{z}_C=0$.  $x_{12}^{\star}$ therefore corresponds to the intersection between the contour of zero vertical fluid velocity induced by a particle,  $u_z=0$, and the line $z=1$ as shown on Fig.\ \ref{fig:fig2}d. If $x_{12}>x_{12}^{\star}$ then $\dot{z}_{12}>0$ and the particles cannot perform the leapfrog motion, as can be intuited by superimposing the flow of two rotlets side by side  (Fig.\ \ref{fig:fig2}a and d). Below a critical value $\xi_c=1$, $\dot{z}_{12}$ is always negative at $z=1$, which implies that there is no critical point $x_{12}^{\star}$ ($u_z=0$ and $z=1$ never meet) and thus that all the trajectories are periodic regardless of the initial separation (Fig.\ \ref{fig:fig2}b-d). The viscosity ratio therefore controls the existence of periodic orbits.
The viscosity ratio also sets the speed and direction of these periodic motions. 
In the limit $\xi=0$ the backward motion is fastest and the height of the system remains constant ($\dot{z}_C=0$) (Fig.\ \ref{fig:fig2}c, left panel). 
Otherwise the direction of motion depends on the relative position of the particles so that, given an initial configuration,  there is always a threshold value for which the pair orbits in place (e.g.\ $\xi = 3.66$ for $x_{12}^0=4$ in Fig.\ \ref{fig:fig2}c, middle panel).
In the limit $\xi=+\infty$, the pair always moves forward: $\dot{x}_{C} = 6z_C x_{12}^2/R^5_{12}>0$ (Fig.\ \ref{fig:fig2}c, right panel). 
\\
 \textit{\textbf{Collective dynamics.}--} 
 We further explore the dynamics at the collective level where large collections of particles interact. In the rigid-wall limit ($\xi\rightarrow\infty$) the suspension exhibits a cascade of events: an initially uniform strip of particles forms a dense front which is subject to a transverse fingering instability, from which the fingertips then detach to generate dense motile structures. As shown in previous work,  the wavelength at the onset of the instability is set by the mean height of the front \cite{Driscoll2017, Delmotte2017b}.
We investigate the effect of the viscosity ratio on this rich collective dynamics using  3D Stokesian dynamics simulations that include both hydrodynamic and steric interactions between tens of thousands of microrollers above the interface \cite{SI}.
Fig.\ \ref{fig:fig3}a shows the time evolution of 10,000 microrollers initially lying on a monolayer with area fraction $\phi=0.65$, at a given height $z_C^0/a=1.2$, for various viscosity ratios $\xi \in \{0, 0.5, 1.5, 5, +\infty\}$ (see SI Movies 1-2).  
In the case of a no-slip wall, $\xi=+\infty$, the strips evolves as described above. The rollers inside the fingertips perform a treadmill motion that follows the clockwise rotation of the torque. A few particles are occasionally shed from the front and lifted up by flow of the rotating structures. When $\xi=5$ the suspension behaves similarly with a slower forward motion. 
At a critical value $\xi \approx 1.5$, the suspension treadmills without translating, but the transverse instability still occurs in place, leading to a lateral separation of the particles.
Below that threshold, the suspension self-assembles into a long roll treadmilling clockwise but moving backwards. When $\xi=0.5$ the roll is wavy due to the transverse instability, but evolves at a significantly slower rate, and sporadically sheds particles from the rear.
In the free-surface limit ($\xi=0$), the transverse instability is suppressed and the roll remains straight and stable. It does not shed any particle because, as shown in the previous section, the  streamlines of the flow induced by the microrollers are closed regardless of the separation distance (Fig.\ \ref{fig:fig2}b, bottom panel). 
Finally, we note that the mean height of the particles in the fingertips slightly increases with $\xi$ (Fig. \ref{fig:fig3}a, bottom panel), which in turn increases the wavelength of the transverse instability. This is due to the upward advective flows ahead of the particles that get stronger with $\xi$ (see (Fig.\ \ref{fig:fig2}b) and allow the particles behind to lift the leading front where the instability occurs. \\
In order to better understand the effect of $\xi$ on the transverse  instability, we use a simple model consisting of two continuous lines of rotlets separated by a distance $d$ in a plane parallel to the interface at a constant height $h$. The lines are aligned along the $y$-axis with rotlet density $\rho_i(y,t)$ and position $x_i(y,t), \, i=1,2$.  The model is governed by the equations of motion of each line together with the conservation of rotlets along their length:
\eqn{
\begin{array}{ll}
\dfrac{\partial x_{i}(y,t)}{\partial t} =& u_x(x_{i}(y,t),y)\\
\dfrac{\partial \rho_{i}(y,t)}{\partial t} =& -\dfrac{\partial [ \rho_{i}(y,t)u_y(x_{i}(y,t),y) ]}{\partial y}\\
\end{array},\,i=1,2,
\label{eq:conserv_lines}
}
where length and time have been rescaled with $l_c=h$ and $t_c$  defined above.
The velocity $(u_x,u_y)$ in the RHS arises from the long-ranged hydrodynamic interactions between the microrollers of each line, e.g.\   
\eqn{
u_x(\underset{\bx}{\underbrace{x_i,y}}) =  \sum_{j=1}^2\int_{-\infty}^{+\infty}\mu_x^{u\tau}(\underset{\bx-\bx'}{\underbrace{x_i-x_j,y-y'}})\rho_j(y')\dd y',
\label{eq:conv_prod1}
} 
where $\mu_x^{u\tau}(\bx-\bx')$ is the $x$-component of the operator $\bmu^{u\tau}(\bx-\bx') = 1/2 \pare{\nabla_{\bx'}\times\bG(\bx-\bx')}\cdot\be_y$  that computes the fluid velocity at position $\bx$ induced by a rotlet, directed along $\be_y$, at position $\bx'$ 
\cite{SI}. As shown in previous work, this model contains all the essential ingredients to faithfully reproduce the fingering instability above a no-slip wall \cite{Delmotte2017b} and naturally extends to fluid-fluid interfaces by using the linear combination of Green's functions in \eqref{eq:Green}. The base state of this system corresponds to two straight lines with uniform rotlet densities ($\rho_i(y,t) = \rho_0$)  translating at a steady speed, so that their position are constant in the moving frame ($x_1(y,t)=0$, $x_2(y,t)=d$). 
After carrying a linear stability analysis of the system around the base state, we obtain an analytical expression for the growth rate $\sigma(k,\xi)$ \cite{SI}, plotted on Fig.\ \ref{fig:fig3}b. 
First, we notice that an identical fastest growing mode is selected for all values of $\xi$, as seen by the clear bump at $\lambda_{\max} = 2\pi/k_{\max} = 23$, except for $\xi=0$ for which there is no clear selection. \rev{The existence of a plateau at large $k$ is specific to the two-lines model, and its dependence on $\xi$ is explained in the SI \cite{SI}.}  Second the corresponding growth rate, $\sigma_{\max}$, increases with $\xi$.
The increase of $\sigma_{\max}$ is quantified on Fig.\ \ref{fig:fig3}c, and shows a  plateau at large $\xi$. A detailed analysis of the two-line model shows that the growth rate of the transverse instability is proportional to the transverse flow $u_y$ in \eqref{eq:conserv_lines}. Since the microrollers do not induce transverse flows above a free surface ($u^{FS}_y=0$ \cite{SI}), $\sigma_{\max}$  scales as $\xi/(\xi+1)$, which is the prefactor of the no-slip wall contribution $\bG^W$ in \eqref{eq:Green}  (see inset of Fig.\ \ref{fig:fig3}c). 
Overall the two-line model highlights the crucial role of these transverse flows and exhibits a good agreement with the simulations in the sense that, for a given particle height, $\xi$ only changes the growth rate of the instability without affecting much the dominant wavelength. However since the height of the particles is kept constant in the model, it cannot capture the slight increase of the wavelength with $\xi$ that is observed on Fig.\ \ref{fig:fig3}a.\\
The nature of the interface does not only affect the direction of motion of the suspension and the growth of the transverse instability, but also the shape and structure of the emerging clusters.
  \begin{figure}[t]
    \centering
    \includegraphics[width=0.7\linewidth]{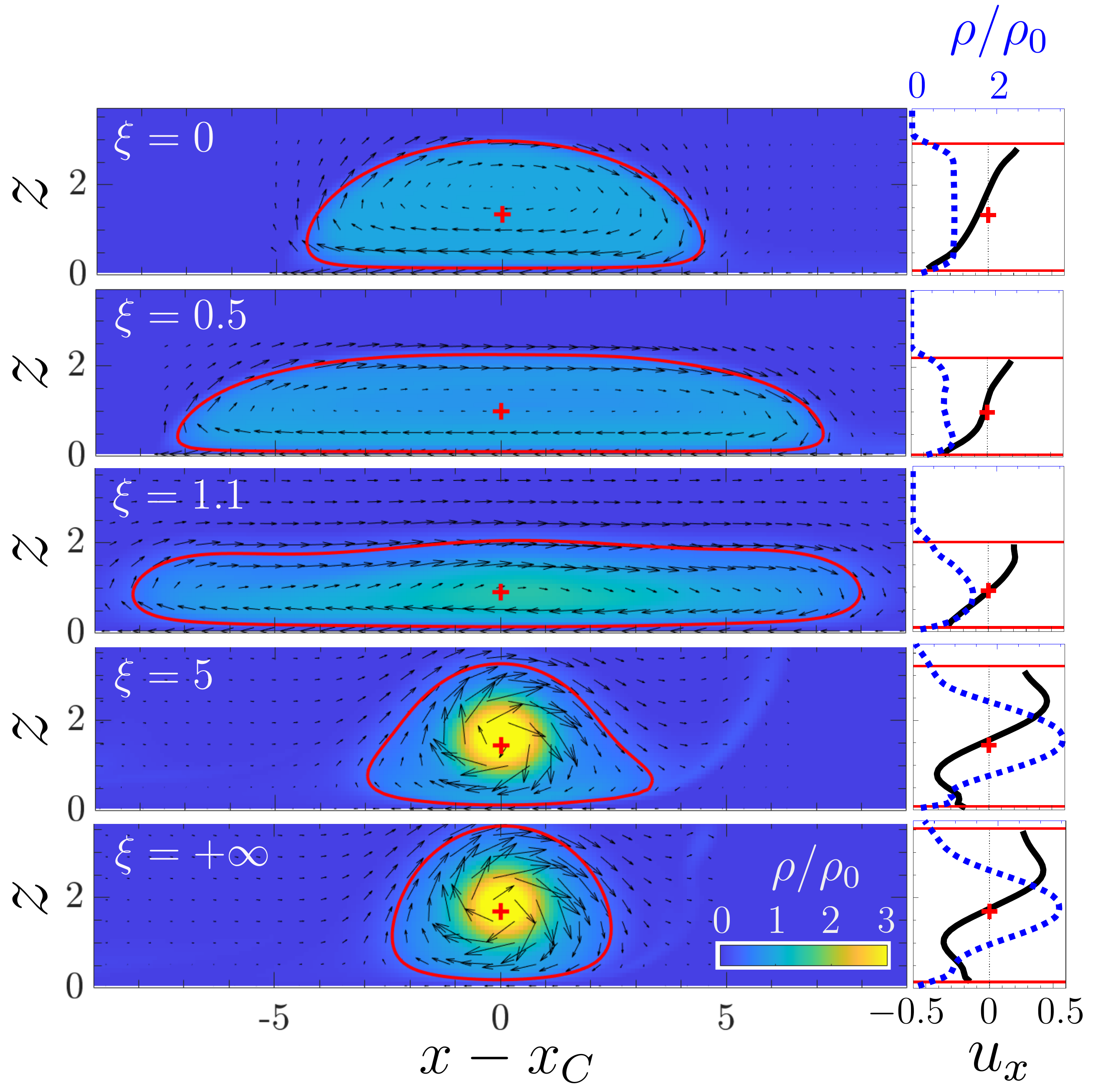}    
     \caption{Particle density distribution $\rho(\bx,t)$ obtained from \eqref{eq:nonlocal_pde} at $t=1418$ (quasi-steady state) in the $(x,z)$-plane for various $\xi \in \{0, 0.5, 1.5, 5, +\infty\}$. Black arrows: flow field. Solid red line: $\rho/\rho_0 = 0.4$. Red cross: position of the COM $(x_C,z_C)$. 
     Side panels: density (dotted blue line) and velocity (solid black line) profiles at $x=x_C$.
     } 
    \label{fig:fig4}
\end{figure}
We study these changes from a macroscopic point of view, with a mean field description of the density of microrollers $\rho(\bx,t)$ in the $(x,z)$-plane perpendicular to the interface, denoted $\mc{P}$.
$\rho(\bx,t)$ obeys a conservation equation  
\eqn{
\frac{\partial \rho}{\partial t} +  \nabla\cdot\pare{\bu\rho} = 0.
\label{eq:nonlocal_pde}
} 
The velocities in the flux term arise from the long-ranged hydrodynamic interactions between microrollers in the plane $\mc{P}$, e.g. $u_x(\bx,t) =  \int_{\mc{P}}\mu_x^{u\tau}(\bx-\bx')\rho(\bx',t)\dd\bx'$.
The nonlocal equation \eqref{eq:nonlocal_pde} is solved numerically \cite{SI} and the rollers are initially uniformly distributed ($\rho=\rho_0$) over a thin strip of aspect ratio \rev{$\Gamma = L/H = 9.4$} near the interface. 
 After some time, the system reaches a quasi-steady state where a main cluster emerges and translates at a steady speed (see SI Movies 3-7). Fig.\ \ref{fig:fig4} shows a snapshot of the microroller density at the same dimensionless time in the quasi-permanent regime for various values of $\xi$ together with the density $\rho(x_C,z)$ and velocity profiles $u_x(x_C,z)$ at the central cross-section of the cluster. 
The cluster is delimited by the iso-value $\rho = 0.4\rho_0$ (red line).
As in the discrete particle simulations, the velocity of the cluster is fastest and backwards for $\xi=0$, vanishes around $\xi\approx 1.1$, where the cluster treadmills in place, and increases forward beyond.
The shape of the cluster evolves from a near-hemisphere of aspect ratio \rev{$\Gamma =3.1$} at $\xi = 0$, to a ``pancake" shape (\rev{$\Gamma = 8.4$}) at $\xi=1.1$, and becomes circular when $\xi$ increases further (\rev{$\Gamma = 2$ and $\Gamma = 1.4$} for $\xi=5$ and $\xi=+\infty$). 
In addition to the shape, $\xi$ controls the particle distribution inside the cluster. The distribution is uniform  at $\xi=0$ ($\rho \approx \rho_0$) and becomes more peaked as $\xi$ increases, with a maximum $\rho_{\max}\approx 3.8\rho_0$ reached at $\xi=+\infty$.
These changes in concentration, together with the boundary condition at the interface underneath, determine the velocity profile within the cluster. At $\xi=0$, the treadmilling motion is fastest near the bottom interface, where the slip condition allows for large velocities. As $\xi$ increases, the particles concentrate at the center and  the no-slip condition becomes stronger, which shifts the maximum velocity upwards and generates a rigid-body motion near the core.\\ 
 \textit{\textbf{Discussion.}--} 
In this Letter, we have shown that the viscosity ratio across interfaces controls both the microscopic and macroscopic response  of an active suspension. These fundamental findings are also of technological importance. Our results suggest that the strategies for particle transport and fluid pumping initially developed with microrollers above solid walls can readily be extended to a variety of biological and microfluidic environments with different type of surfaces. \rev{Our results could even generalize to solid boundaries with a finite slip length \cite{Lauga2005}, which  are found in many experimental systems, such as hydrophobic surfaces.}
In this work we considered particles lying \textit{above} the interface, i.e.\ with large contact angles.  In some cases the particles wet the outer fluid and get adsorbed at the interface.  Adsorbed active particle layers can be used as active surfactants to modulate interfacial properties in emulsions or films, or to pump and mix flows in the surrounding fluids \cite{Fei2017,Fei2020}. Up to now,  the hydrodynamics of torque-driven particles straddling fluid-fluid interfaces is still not well-understood \cite{Maldarelli2022} and should deserve more attention given their exciting applications. \\

I thank M. Driscoll, A. Donev and S. Michelin for their critical reading of the manuscript as well as for stimulating discussions. I acknowledge support from the French National Research Agency (ANR), under award ANR- 20-CE30-0006. I also thank the NVIDIA Academic Partnership program for providing GPU hardware for performing some of the simulations reported here.





\nocite{Usabiaga2016b,Bell1988,May2011}
\bibliography{Biblio}

\end{document}



\title{Supplementary Materials\\ 
to\\
Viscosity ratio across interfaces controls the stability and self-assembly of microrollers}

\author{Blaise Delmotte}
\email{blaise.delmotte@cnrs.fr}
 \affiliation{LadHyX, CNRS, Ecole Polytechnique, Institut Polytechnique de Paris, 91120 Palaiseau, France}

\maketitle

\section{Far-field approximation of the self-induced velocity of a single microroller above a fluid-fluid interface}
The flow induced by a particle located at $\bx'$ and driven by a torque $\btau$ is given by
\eqn{
\bu(\bx) = \fr{1}{2}\corchete{\nabla_{\bx'}\times\bG(\bx-\bx')}\cdot\btau
}
where $\bG(\bx-\bx') =1/(\xi+1)\bG^{FS}(\bx-\bx') + \xi/(\xi+1)\bG^W(\bx-\bx')$ is the Green's function of the domain (Eq.\ (1) in the main text).
The self-induced velocity $\bv_0$ of the particle due to its own flow is given by the Faxen formula 
\eqn{
\bv_0 = \left.\pare{1+\fr{a^2}{6}\nabla^2}\bu(x)\right|_{\bx=\bx'}.
}
Using the analytical expressions of $\bG^{FS}$  and $\bG^W$ provided in \cite{Lee1979} and  \cite{Blake1974} respectively, the self-induced response to an external torque aligned along the $y$-axis,  $\btau=\tau\be_y$,  is given by  $\bv_0=v_0\be_x$ where 
\eqn{
v_0 =  \frac{\tau}{32\pi\eta a^2} \pare{\frac{a}{h}}^2\left[-\frac{1}{\xi +1} + \frac{\xi}{\xi+1}\frac{1}{2}\pare{\frac{a}{h}}^2\right],
}
which leads to Eq.\ (2) in the main text after normalizing by $\tau/8\pi\eta a^2$. Note that the self-induced velocity is always negative in the point-particle limit $(a/h \rightarrow 0$)
\eqn{
\lim_{a/h\rightarrow 0} v_0 =  -\frac{\tau}{32\pi\eta} \frac{1}{\xi +1} \frac{1}{h^2}\le 0
\label{eq:self_ind_rotlet}
}

\section{Dynamical system approach for the two point-torques (rotlets)}
\subsection{Equations of motion}
Consider two point particles with positions $\bx=(x_i,z_i),\, i=1,2$ driven by a torque $\btau$ in the ($x,z$)-plane above a fluid-fluid interface.
Since the particles have no size, the velocity of particle $1$ due to the torque on particle $2$ is equal to the fluid velocity at the particle position $\bx_1$: 
\eqn{\bv_1 = \bu_2(\bx=\bx_1)=\fr{1}{2}\left.\corchete{\nabla_{\bx'}\times\bG(\bx_1-\bx')}\right|_{\bx'=\bx_2}\cdot\btau}
From this formula, and by adding the self-induced contribution \eqref{eq:self_ind_rotlet}, one obtains simple analytical expressions for the particle velocities $\bv_i = (\dot{x}_i,\dot{z}_i),\, i=1,2$. Using the translational invariance of the system along the $x$-direction, one can reduce the system to three degrees of freedom, $(x_{12}=x_1-x_2,z_{12}=z_1-z_2,z_C=(z_1+z_2)/2)$, which leads to Eq.\ (3) in the main text. \rev{Note that the mean horizontal position $x_C = (x_1+x_2)/2$ is omitted in Eq.\ (3). Due to the translational invariance of the system along the $x$-axis, $x_C$ decouples from the other degrees of freedom in the sense that it does not affect their evolution in Eq.\ (3) (the contrary is not true, $\dot{x}_C$ does depend on $x_{12},\, z_{12}$ and $z_C$). As a result $x_C$ does not affect the existence, and the values, of the critical points that separate periodic and non-periodic orbits (see section below).}

\subsection{Critical saddle points}
 \begin{figure}[t]
    \centering
    \includegraphics[width=0.6\columnwidth]{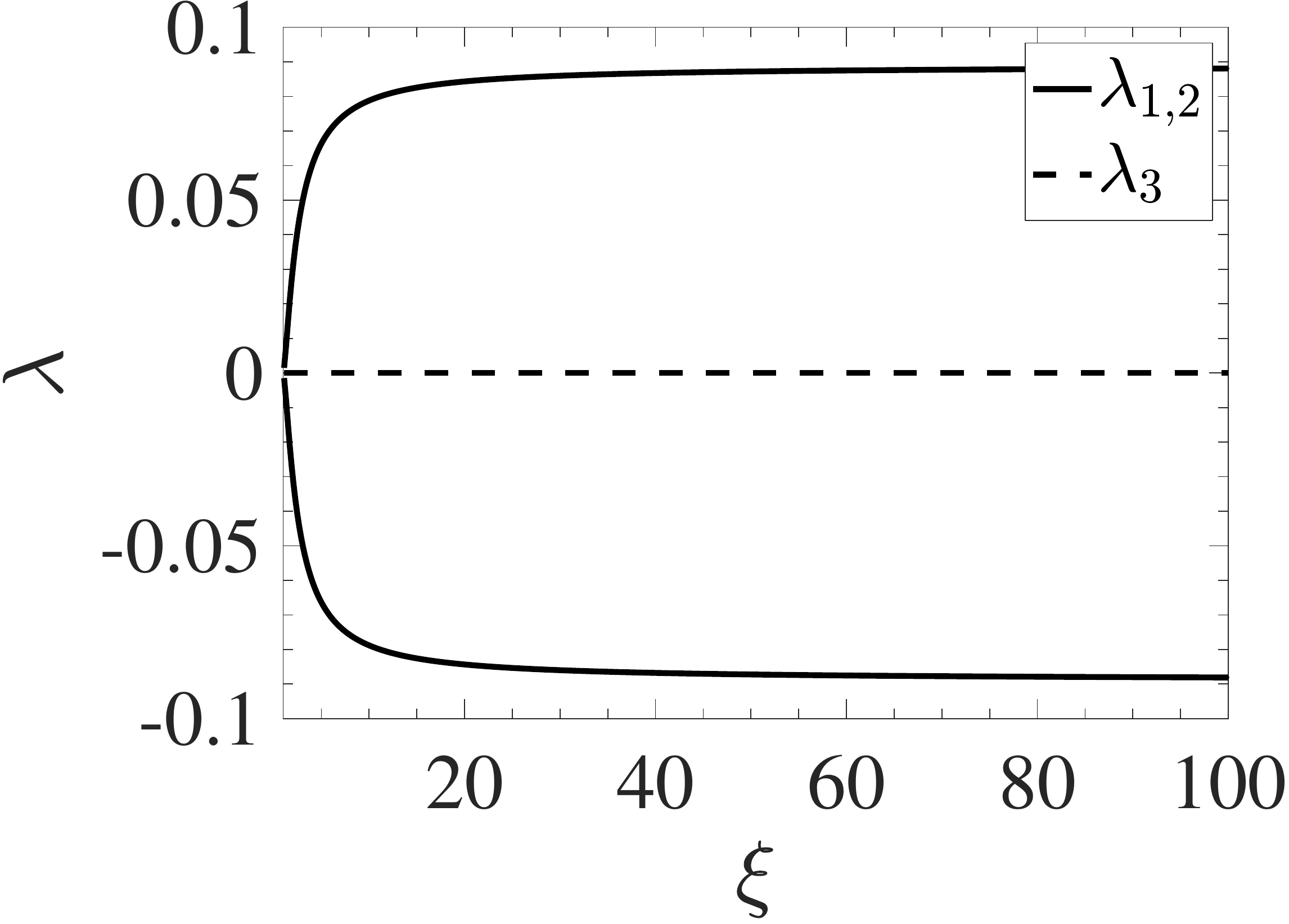}    
    \caption{Eigenvalues of $\left.\nabla_{\bx} \mc{G}(\bx)\right|_{\bx^{\star}_{\pm}}$ as a function of the viscosity ratio $\xi \in [1,100]$ } 
    \label{fig:EV}
\end{figure}
The critical points of the dynamical system are found by solving 
\eqn{
\left[\begin{array}{c}
\dot{x}_{12}\\
\dot{z}_{12}\\
\dot{z}_C
\end{array}\right]  =  \left[\begin{array}{c} 2z_{12}\left(\frac{1}{r_{12}^3} - \fr{\xi}{\xi+1} \frac{2x_{12}^2-4z_C^2}{R_{12}^5}\right) -\frac{1}{4}\fr{1}{\xi+1}\pare{\fr{1}{(z_C+z_{12}/2)^{2}}-\fr{1}{(z_C-z_{12}/2)^{2}}}\\
 2x_{12}\left(\frac{1}{R_{12}^3} - \frac{1}{r_{12}^3} +  \fr{\xi}{\xi+1}\frac{12 z_C^2}{R_{12}^5}\right) \\
\fr{\xi}{\xi+1} \frac{6 x_{12}z_{12}z_C}{R_{12}^5}
 \end{array}\right] = \mc{G}(\bx) = \bzero
}
where $\mc{G}(\bx)$ is the vector field that describes the body dynamics in phase space.
Their coordinate is given by $\bx^{\star}_{\pm} = ( x_{12}=\pm x_{12}^{\star},z_{12}=0,z_C=1)$, where $x_{12}^{\star}\neq 0$ is a zero of  $\dot{z}_{12}(x_{12},0,1,\xi)$, which is found by solving 
\eqn{
x_{12}^3\left(x_{12}^2+4+12\frac{\xi}{\xi+1}\right) - \left(x_{12}^2+4\right)^{5/2} = 0.
}
In order to characterize these fixed points, we compute the eigenvalues of the Jacobian matrix of the system evaluated at $\bx^{\star}$ ($\left.\nabla_{\bx} \mc{G}(\bx)\right|_{\bx^{\star}_{\pm}}$). 
Fig.\ \ref{fig:EV} shows these eigenvalues as a function of $\xi \in [1,100]$ (remember that the critical points do not exist for $\xi<1$, see main text). Two eigenvalues are real with opposite signs and the third one is zero, which clearly shows that these critical points are saddle points and that the dynamics lies on a two-dimensional manifold in the vicinity of $\bx^{\star}_{\pm}$ in phase space. These two saddle points with symmetric positions delimit a separatrix between periodic and non-periodic orbits. Since $x_{12}^{\star}$ decreases when $\xi$ increases, the size of the domain enclosed by the separatrix decreases as well.
Note that the magnitude of the eigenvalues increases with $\xi$,  which indicates an acceleration of the relative motion between particles with $\xi$ in the vicinity of the fixed points.

\section{Details on the large scale 3D Stokesian dynamics simulations}
Consider a collection of $N$ microrollers with radius $a$ and positions $\bX=\{\bx_1,\dots,\bx_N\}$ subject external forces $\bF=\{\bbf_1,\dots,\bbf_N\}$ and torques $\bT=\{\btau_1,\dots,\btau_N\}$, where $\btau_1=\dots=\btau_N = \btau =  \tau\be_y$.
The equation of motion of the particles follow the mobility problem:
\eqn{
\dot{\bX} = \bV = \bM^{VF}\cdot\bF + \bM^{VT}\cdot\bT
\label{eq:mob_prob}
}
where $\bM^{VF}$ and $\bM^{VT}$ are the mobility matrices that relate the particle velocities to their external forces and torques respectively. Similarly to Section I, they are obtained by applying the Faxen laws to the flow induced by the particles above the  interface, see \cite{Swan2007,Driscoll2017} for more details on their calculation. This modelling includes only leading-order corrections for the finite size of the particle to limit the computational cost required to simulate large numbers ($O(10^4)$) of particles. Even though this low-resolution model overestimates the particle mobility, our previous work shows that it reproduces qualitatively well the experimental results above a no-slip boundary  \cite{Driscoll2017,Usabiaga2016b,Delmotte2017}.

The external force on each microroller $\bbf_i$ is the resultant of a short-ranged repulsion between particles $\bbf_P$, a short range repulsion with the interface $\bbf_I$ to avoid overlaps, and gravity $\bbf_G$, so that  $\bbf_i = \bbf_P + \bbf_I + \bbf_G$.
The short range repulsive forces $\bbf_P$ and $\bbf_I$ derive from an exponentially decaying repulsive potential of the form \cite{Usabiaga2016b}
\begin{equation}
    U(r) = \begin{cases}
U_0 (1+\frac{d-r}{b}) & \mbox{if } r<d,\\
U_0\exp(\frac{d-r}{b})  & \mbox{if } r\ge d.\\
\end{cases}
\label{eq:potential}
\end{equation}
For particle-particle interactions, $r$ is the center-to-center distance and $d=2a$. For particle-interface interactions, $r$ is the height of the particle center and $d=a$. The energy scale $U_0$ and interaction range $b$ control the strength and decay of the potential respectively. 
We found that taking $b=0.1a$,  $U_0 = \tau/2$ for particle-interface interactions and $b=0.1a$, $U_0 = \tau/26$ for particle-particle interactions prevents overlaps while keeping close contact.\\
In the large scale particle simulations, a small amount of gravity is added: $\bbf_G = -mg \be_z$, with $mg=0.0015\tau/a$. 
This is done because, in the absence of gravity, the front particles are lifted up quite high by the rear particles when $\xi=+\infty$. As a result, the transverse instability would take a long time to develop and, because it is proportional to the particle height \cite{Driscoll2017,Usabiaga2016b,Delmotte2017b}, the wavelength of the instability would be large and becomes similar to the domain size. The addition of a small gravitational force slightly reduces the height of the front particles and therefore increases the growth rate of the instability and allows to have several wavelengths along the domain. However, the effects of gravity remain negligible compared to the torque induced flows ($\tau\gg mga$) and we have carefully checked that it did not change the qualitative behaviour of the suspension for all the values of $\xi$ simulated.\\
Particle trajectories \eqref{eq:mob_prob} are integrated with the two-step Adams–Bashforth  scheme.  
Mobility-vector products and steric interactions are computed with PyCUDA on an Nvidia Titan V GPU. The typical simulation time is 30 min for 4,000 time iterations with 10,000 particles.

\section{Two-lines model}
This section describes the two-line model and the details of the linear stability analysis. This model has already been used to describe the transverse instability of micro-rollers above a no-slip wall \cite{Delmotte2017b}. The derivation and analysis below are a direct extension of this work, where only the Green's function has been changed. For the sake of  clarity we recall the governing equations and the main steps of the stability analysis.

 \begin{figure}[h!]
    \centering
    \includegraphics[width=0.6\columnwidth]{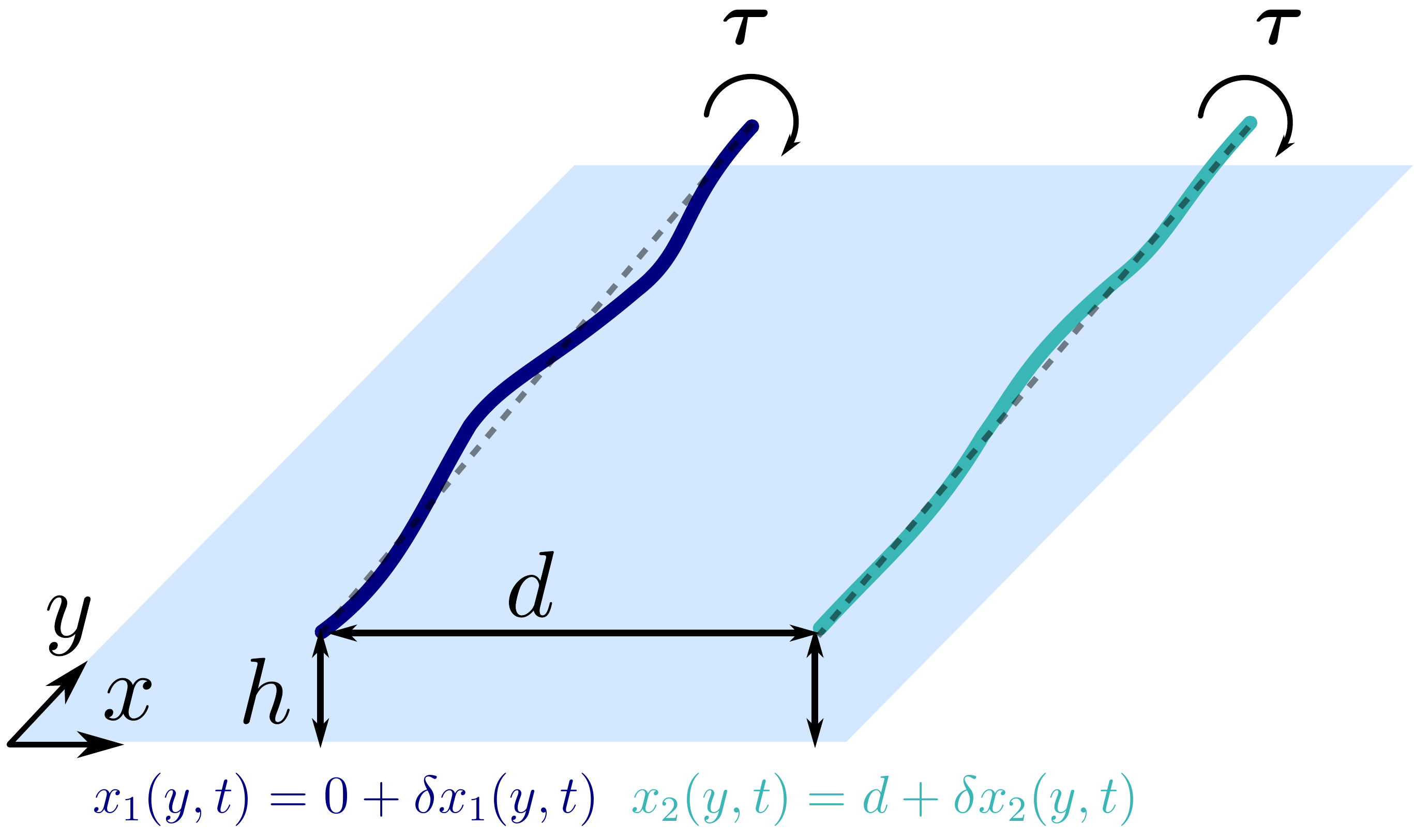}
    \caption{Sketch of the two infinite lines. The rotlet densities are rotating with a constant 
torque, $\tau$, about the $y$-axis and interact hydrodynamically in the $xy$-plane at a height $h$ above the floor.} 
    \label{fig:sketch_lines}
\end{figure}

\subsection{Governing equations}
\begin{figure}[h!]
    \centering
    \includegraphics[width=0.6\columnwidth]{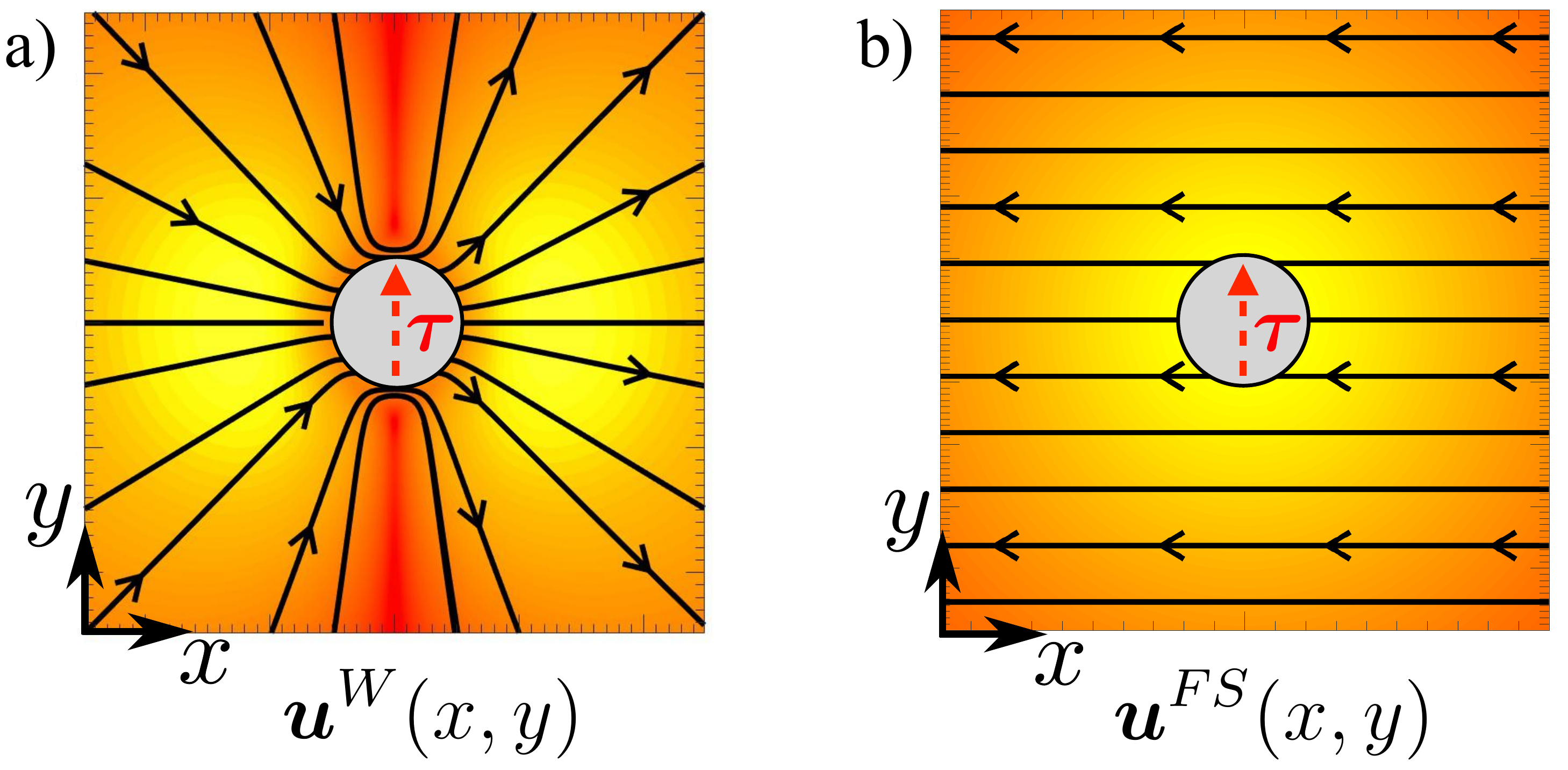}
    \caption{Flow induced by a microroller in the plane parallel to the surface. a) No-slip wall (W), b) Free-surface (FS).} 
    \label{fig:flow_slip_no_slip}
\end{figure}

Consider two infinite lines of rotlets, labelled $\mc{L}_1$, at the rear, and $\mc{L}_2$, at the front, rotating around the $y$ axis at a fixed height $h$ in a plane parallel to the fluid-fluid interface (see Fig.\ \ref{fig:sketch_lines}).  We denote  $x_1(y,t)$ and $x_2(y,t)$ their position and  $\rho_1(y,t) \equiv \rho(x_1(y,t),y,t)$ and $\rho_2(y,t)\equiv \rho(x_2(y,t),y,t)$ their rotlet densities. 
Here, we neglect out of plane motion in the $z$-direction and only consider the velocities in the $xy$-plane. Although out of plane motion is seen in 3D simulations, our previous work \cite{Driscoll2017,Delmotte2017b} shows that the fingering instability still occurs when considering  particles confined to a plane above the surface.\\
The model is governed by the equations of motion of each line together with the conservation of rotlets along their length:
\eqn{
\begin{array}{ll}
\dfrac{\partial x_{i}(y,t)}{\partial t} =& u_x(x_{i}(y,t),y)\\
\dfrac{\partial \rho_{i}(y,t)}{\partial t} =& -\dfrac{\partial [ \rho_{i}(y,t)u_y(x_{i}(y,t),y) ]}{\partial y}\\
\end{array},\,i=1,2,
\label{eq:conserv_lines}
}
where length and time have been rescaled with $l_c=h$ and $t_c = 8\pi\eta l_c^3/\tau$  respectively.  These  four PDEs  are nonlinear and nonlocal. 
The velocity at a given point $(x_i,y)$ along a line $\mc{L}_i$ is given by the sum of the flows induced by each line, 
\eqn{
u_x(x_i,y) =  \sum_{j=1}^2\int_{-\infty}^{+\infty}\mu_x^{u\tau}(x_i-x_j,y-y')\rho_j(y')\dd y',\\
u_y(x_i,y) =  \sum_{j=1}^2\int_{-\infty}^{+\infty}\mu_y^{u\tau}(x_i-x_j,y-y')\rho_j(y')\dd y',
\label{eq:functionals}
}
where  $\bmu^{u\tau}(\bx-\bx')$ is the operator which provides the fluid velocity in the plane at position $\bx = (x,y,h)$ induced by a rotlet at position $\bx'=(x',y',h)$. It is given by 
\eqn{
  \mu_x^{u\tau}(x-x',y-y') &= 1/2 \corchete{\pare{\nabla_{\bx'}\times\bG(x-x',y-y';h)}\cdot\be_y}\cdot\be_x \\
&= \fr{\xi}{\xi+1} 6h^3\frac{(x-x')^2}{R^{5}} - \fr{1}{\xi+1} 2\dfrac{h^3}{R^{3}}\\
&=\fr{\xi}{\xi+1}u_x^W + \fr{1}{\xi+1}u_x^{FS}\\
  \mu_y^{u\tau}(x-x',y-y') &= 1/2 \corchete{\pare{\nabla_{\bx'}\times\bG(x-x',y-y';h)}\cdot\be_y}\cdot\be_y \\
&= \fr{\xi}{\xi+1} 6h^3\frac{(x-x')(y-y')}{R^{5}} +\fr{1}{\xi+1} \times 0\\
&=\fr{\xi}{\xi+1}u_y^W + \fr{1}{\xi+1}u_y^{FS}
}
 where $R=\left[(x-x')^2 + (y-y')^2 + 4h^2\right]^{1/2}$. As shown on Figure \ref{fig:flow_slip_no_slip} the free-surface contribution has no transverse flow: $u^{FS}_y=0$.  

\subsection{Linear stability analysis}
 The base state of this system corresponds to two straight lines with uniform rotlet densities ($\rho_i(y,t) = \rho_0$)  translating at a steady speed, so that their position are constant in the moving frame ($x_1(y,t)=0$, $x_2(y,t)=d$), where $d$ is the initial distance between the two lines.
 
 We perturb the system about the base state 
\begin{eqnarray}
 x_1(y,t)  =  0 + \delta x_1(y,t)\\ 
 x_2(y,t)  =  d + \delta x_2(y,t),\\
 \rho_1(y,t)  =  \rho_{0} + \delta \rho_1(y,t)\\
 \rho_2(y,t)  =  \rho_{0} +  \delta \rho_2(y,t),
\end{eqnarray}
where $\delta x_1, \delta x_2\ll \min(d,h)$ and  $\delta \rho_1, \delta \rho_2 \ll \rho_0$.

After Taylor expanding the functionals in 
 Eqs.\ \eqref{eq:conserv_lines}-\eqref{eq:functionals} we obtain the linearized governing equations
\begin{eqnarray}
\dfrac{\partial \delta x_1}{\partial t} &=& \rho_0 \mathcal{G}_0\delta x_1 +  \rho_0 \mathcal{G}_1*\delta x_2  +  \mu_x^{u\tau}(0-d,\cdot)*\delta\rho_2 \label{eq:adv_linearized}\\
\dfrac{\partial \delta x_2}{\partial t} &=& -\rho_0 \mathcal{G}_0\delta x_2 - \rho_0 \mathcal{G}_1*\delta x_1  +  \mu_x^{u\tau}(0-d,\cdot)*\delta\rho_1\label{eq:adv_linearized2}\\
\dfrac{\partial \delta\rho_1}{\partial t} &=& -\rho_0\dfrac{\partial \left[ \left(\rho_0 \mathcal{G}_2*\delta x_1 + \rho_0 \mathcal{G}_3*\delta x_2  +  \mu_y^{u\tau}(0-d,\cdot)*\delta\rho_2 \right)\right]}{\partial y}\\
\dfrac{\partial \delta\rho_2}{\partial t} &=& -\rho_0\dfrac{\partial \left[ \left(\rho_0 \mathcal{G}_2*\delta x_2 + \rho_0 \mathcal{G}_3*\delta x_1  -  \mu_y^{u\tau}(0-d,\cdot)*\delta\rho_1 \right)\right]}{\partial y},\label{eq:continuity2_linearized}
\end{eqnarray}
where we have used the symmetries of $\mu_x^{u\tau}$ and $\mu_y^{u\tau}$. The star ``$*$" denotes the one dimensional convolution product. 
The interaction kernels are given by 
\eqn{
\mathcal{G}_0 =& \left.\dfrac{\partial}{\partial x_1} \left[\int\limits_{-\infty}^{+\infty} \mu_x^{u\tau}\left(x_1-d,y-y'\right)  dy'\right]\right\rvert_{x_1=0}  \label{eq:G0}\\
\mathcal{G}_1(y-y') =& \left.\dfrac{\partial \mu_x^{u\tau}(0-x_2,y-y')}{\partial x_2}\right\rvert_{x_2 = d} ,\\
\mathcal{G}_2(y-y') =& \left.\dfrac{\partial \mu_y^{u\tau}(0-x_1,y-y')}{\partial x_1}\right\rvert_{x_1 = 0}, \label{eq:G2}\\
\mathcal{G}_3(y-y') =& \left.\dfrac{\partial \mu_y^{u\tau}(0-x_2,y-y')}{\partial x_2}\right\rvert_{x_2 = d}.
}
Note that $\mathcal{G}_0$ is a constant which depends only on the geometric parameters of the problem $d$, $h$. 

Looking for  periodic solutions of Eqs.\ \eqref{eq:adv_linearized}-\eqref{eq:continuity2_linearized} of the form
$$\mathbf{u} = (\delta x_1, \delta x_2, \delta \rho_1, \delta\rho_2) = \tilde{\mathbf{u}} \; e^{iky+\sigma t},$$
where the wavenumber $k = 2\pi/\lambda$ and
$$\tilde{\mathbf{u}} = (\delta \tilde{x}_1, \delta \tilde{x}_2, \delta \tilde{\rho}_1, \delta \tilde{\rho}_2),$$
we obtain the following eigenvalue problem
\begin{equation}
  A\tilde{\mathbf{u}} = \sigma \tilde{\mathbf{u}}
  \label{eq:EV_problem}
\end{equation}
where
\begin{equation}
   A=  \left[\begin{array}{cccc}
\rho_0\mathcal{G}_0 & \rho_{0}\tilde{\mathcal{G}}_1 & 0 & \tilde{\mu}_x^{u\tau}(-d,k)\\
-\rho_{0}\tilde{\mathcal{G}}_1 & -\rho_0\mathcal{G}_0 & \tilde{\mu}_x^{u\tau}(-d,k) & 0\\
-ik\rho_{0}^{2}\tilde{\mathcal{G}}_2 & -ik\rho_{0}^{2}\tilde{\mathcal{G}}_3 & 0 & -ik\rho_{0}\tilde{\mu}_y^{u\tau}(-d,k)\\
-ik\rho_{0}^{2}\tilde{\mathcal{G}}_3 & -ik\rho_{0}^{2}\tilde{\mathcal{G}}_2 & ik\rho_{0}\tilde{\mu}_y^{u\tau}(-d,k) & 0
\end{array}\right], \label{eq:MatA}
\end{equation}
and the tilde denotes the Fourier transform with respect to $y$, e.g.
\eqn{
\tilde{\mu}_x^{u\tau}(-d,k)=\int\limits_{-\infty}^{\infty} \mu_x^{u\tau}(-d,y)\exp(-iky)dy. 
}
For the sake of conciseness, the exact expressions of the entries of $A$ are not reported here. They correspond to linear combinations of the expressions given  in the main text and appendix A of \cite{Delmotte2017b}.
Due to the particular structure of $A$, Eq.\ \eqref{eq:EV_problem}  can be solved analytically. The 
 four solutions $\sigma_1(k,\xi),..,\sigma_4(k,\xi)$ are given by
\begin{eqnarray}
\sigma_{1,2,3,4}(k,\xi) &=& \pm\frac{1}{2}\left[2A_{11}^2 - 2A_{12}^2 + 4A_{32}A_{14}-2A_{34}^2\right. \nonumber\\
& &\pm 2\left(A_{11}^4-2A_{11}^2A_{12}^2+4A_{11}^2A_{14}A_{32}+2A_{11}^2A_{34}^2-8A_{11}A_{14}A_{31}A_{34} +A_{12}^4\right. \nonumber\\
&& \left. \left.-4A_{12}^2A_{14}A_{32}-2A_{12}^2A_{34}^2+8A_{12}A_{14}A_{32}A_{34}+4A_{14}^2A_{31}^2-4A_{14}A_{32}A_{34}^2+A_{34}^4 \right)^{1/2} \right]^{1/2}.\nonumber\\
\label{eq:EVS}
\end{eqnarray}
 The first two eigenvalues are real and of opposite sign $\sigma_1 = -\sigma_2$. The two other are imaginary and conjugate  $\sigma_3 = \bar{\sigma_4}$.
Figure 3b in the main text shows the positive real eigenvalue $\sigma_{1}$, i.e. the unstable one, for $d = 10h$.

\subsection{\rev{Effect of the viscosity ratio on the growth rate at large wavenumbers}}
 \rev{One specificity of the two-lines model, that is not seen in 3d numerical simulations, is the existence of a plateau at large $k$ (see Fig.\ 3b in the main text). This plateau was already identified in the case of a no-slip wall \cite{Delmotte2017}.  This plateau is specific to the two-lines model, which, even though far from the actual system, is the simplest representation we could find to carry out a linear stability analysis analytically. Its value is set by the term $\mc{G}_0$ in eqs. \eqref{eq:adv_linearized}-\eqref{eq:adv_linearized2} and \eqref{eq:G0}: $\lim_{k\rightarrow \infty}\sigma = \rho_0 |\mc{G}_0|$.  This term represents the uniform, unidirectional, advective flow induced by a straight line with a uniform density of rotlets on the other line at a distance $d$. Because it depends on the local distance between the lines, this flow amplifies shape perturbations of the other line at all scales (see Fig.\ 3d in \cite{Delmotte2017} for an example above a no-slip wall). In the case of a fluid-fluid interface, it has two contributions, one from the wall term and one from free-surface term: 
\eqn{
\mc{G}_0 &= \frac{\xi}{\xi+1} \mc{G}_0^W + \frac{1}{\xi+1} \mc{G}_0^{FS}\\
 &= \frac{\xi}{\xi+1} \fr{8}{3} \fr{d (d^2-4h^2)}{(d^2+4h^2)^3} - \frac{1}{\xi+1} \fr{4}{3} \fr{d}{(d^2+4h^2)^2}
} 
where $d$ is the separation distance between the two lines and $h$ their height above the interface. Note that $\mc{G}_0$ does not depend on $k$ and therefore, unlike the other terms in the stability analysis, does not vanish in the large $k$ limit. In the case of a no-slip wall ($\xi = +\infty$), $\mc{G}_0$ is positive because the lines move forward. Above a free-surface ($\xi=0$), $\mc{G}_0$ is negative because the lines move backward. There is therefore a value of the viscosity ratio $\xi_*$ for which these two contributions cancel:
\eqn{
\xi_* = \frac{1}{2}\frac{d^2+4h^2}{d^2-4h^2}.
}
When $\xi=\xi_*$, the plateau at large $k$ vanishes ($\mc{G}_0=0$).
In the case of Fig.\ 3d, I had chosen $d=10h$, which gives $\xi_* = 13/24 = 0.541667$.\\
Fig.\ \ref{fig:gr} below shows the growth rate in the range close to $\xi_*$. We can clearly see that the value of the plateau decreases, vanishes at $\xi_* = 13/24$ and increases again to reach  $\rho_0|\mc{G}_0| =  1.03\times 10^{-4}$ (in dimensionless units) at $\xi=0$.\\
 \begin{figure}[h!]
    \centering
    \includegraphics[width=0.6\columnwidth]{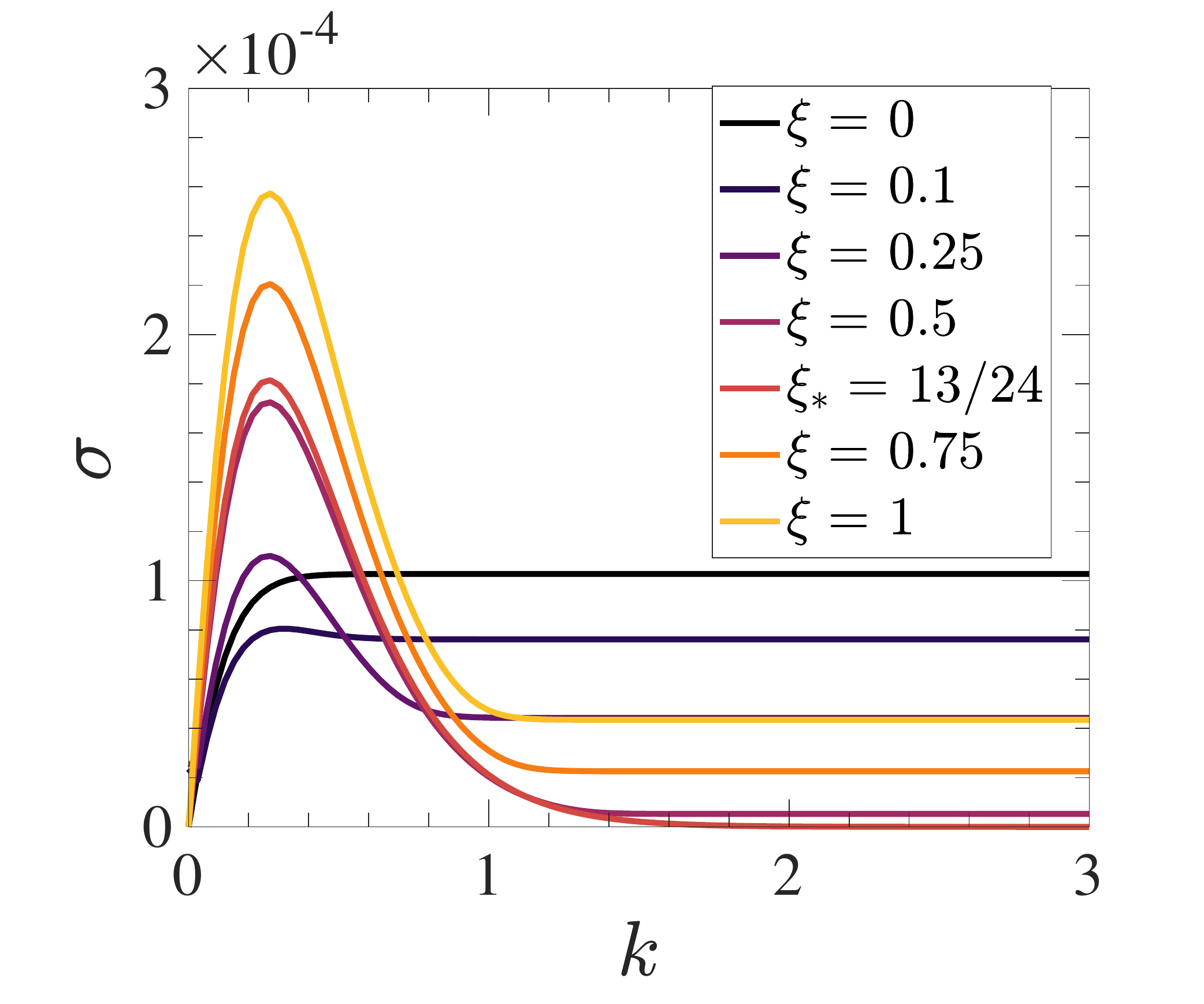}
    \caption{\rev{Growth rate of the two-line model in the range $\xi \in [0;1]$}} 
    \label{fig:gr}
\end{figure}
However, even though this $\mc{G}_0$ term arises naturally in the two-lines model, it does not affect the lengthscale selection of the transverse instability:  if one removes it from the linear stability analysis (see Fig.\ \ref{fig:gr_no_g0}), the fastest growing mode is still preserved, the plateau disappears and the system because fully stable in the free-surface case (as in the 3d Stokesian Dynamics simulation shown on Fig.\ 3a).\\
 \begin{figure}[h!]
    \centering
    \includegraphics[width=0.6\columnwidth]{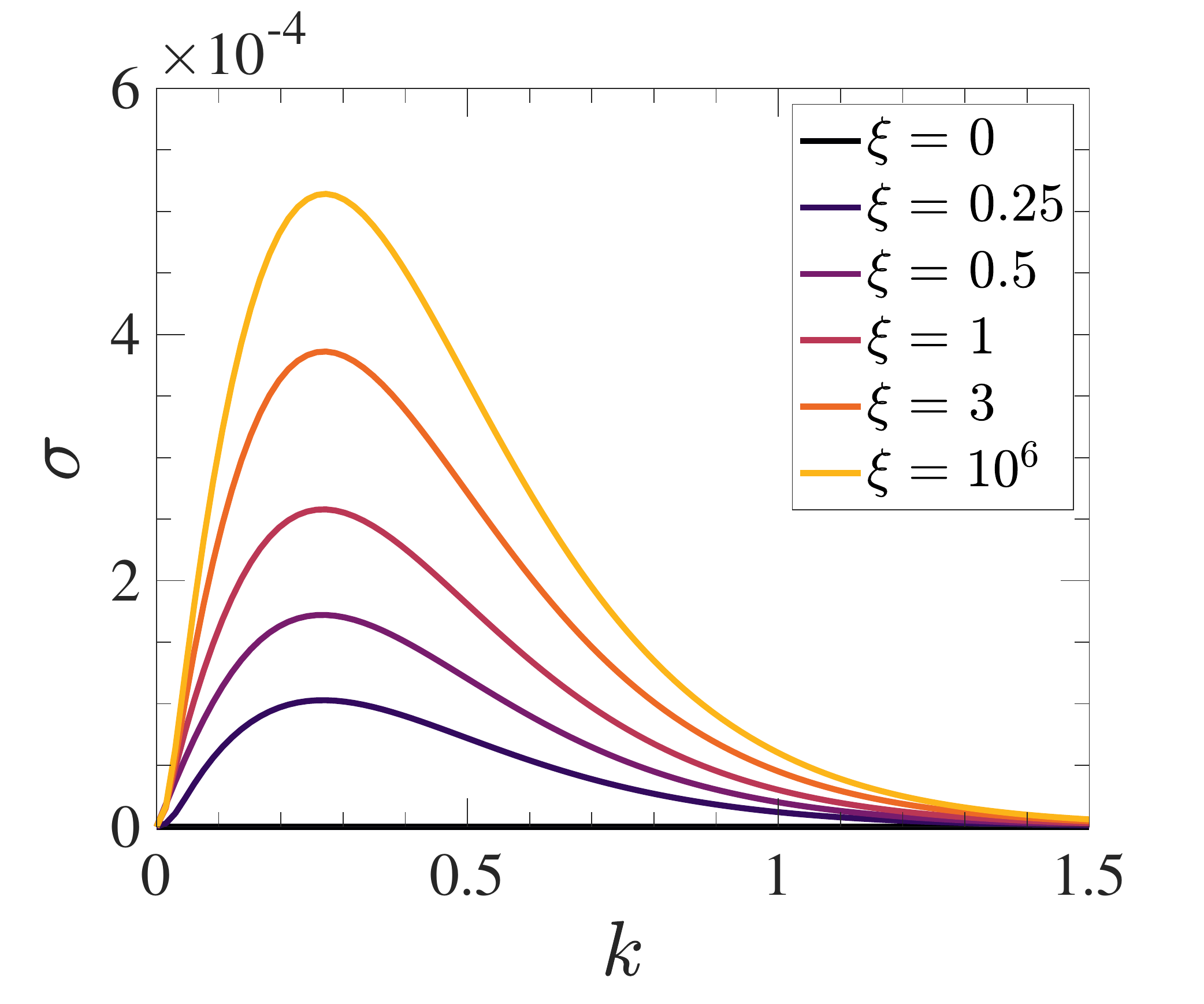}
    \caption{\rev{Growth rate of the two-line model in the range $\xi \in [0;10^6]$ without the term $\mc{G}_0$.}} 
    \label{fig:gr_no_g0}
\end{figure}
}
\section{Details on the mean-field simulations}
The nonlocal mean-field equation Eq.\ (6) in the main text is solved numerically  with a third-order second-time finite volume method optimally designed for hyperbolic scalar conservation equations \cite{Bell1988,May2011}. The rectangular computational domain is discretized with a cartesian grid with cell sizes $\Delta x$ and $\Delta z$ respectively. The domain dimensions $L_x$ and $L_z$ are chosen to ensure that the roller density remains small at the boundaries ($\rho/\rho_0<\epsilon=10^{-4}$) during the simulation time. The time step $\Delta t$ is automatically chosen at each time iteration to ensure that the CFL condition is always satisfied: $\Delta t = 0.9 \min(\Delta x/u_x,\Delta z/u_z)$. 
The convolution products due to hydrodynamic interactions,
\eqn{
u_x(\bx,t) &=  \int_{\mc{P}}\mu_x^{u\tau}(\bx-\bx')\rho(\bx',t)\dd\bx'\\
u_z(\bx,t) &=  \int_{\mc{P}}\mu_z^{u\tau}(\bx-\bx')\rho(\bx',t)\dd\bx',
}
are evaluated with a Riemann sum using PyCUDA on an Nvidia Titan V GPU for efficiency. 
The typical simulation time is 50 min for 9,000 time iterations with $N = N_x \times N_z =  1500 \times 200 = 3\cdot 10^5$ grid cells.\\
Note that corrections for the finite size of the particles could be included in the hydrodynamic interactions in Eq.\ (7) (as done in the particle simulations) with no additional complexity, but these effects are negligible at the macroscopic level where collective effects dominate at large scales.

\newpage
\section{Movie captions}

\begin{itemize}
    \item SI Movie 1, ``Particle\_simulations\_compare\_xi\_top\_view.avi": Top view of the time-evolution of 10,000 microrollers initially uniformly distributed in a monolayer above the interface. Each color, from dark blue to yellow,  corresponds to a different value of $\xi \in \{0,0.5,1.5,5,+\infty\}$ simulated independently with the 3D Stokesian Dynamics approach. 
    \item SI Movie 2, ``Particle\_simulations\_compare\_xi\_side\_view.avi": Side view of the numerical simulations shown in SI Movie 1.
    \item SI Movie 3, ``Continuum\_model\_xi\_0.mp4": numerical simulation of the continuum model above an interface with viscosity ratio $\xi=0$ (free-slip). The rollers are initially uniformly distributed ($\rho/\rho_0=1$) over a thin strip of aspect ratio $\gamma = L/H = 9.4$ near the interface. Colorbar: normalized particle density $\rho/\rho_0$. Black arrow: fluid velocity field. The boundary of the cluster is defined by the green isocontour $\rho/\rho_0=0.4$. The motion along  the $x$-axis is expressed relative to the initial position of the center of mass, $x_C^0 = x_C(t=0)$, in order to show how far the cluster has travelled at the end of the simulation.
    \item SI Movie 4, ``Continuum\_model\_xi\_0\_5.mp4": same as SI Movie 3 but with $\xi = 0.5$.
    \item SI Movie 5, ``Continuum\_model\_xi\_1\_1.mp4": same as SI Movie 3 but with $\xi = 1.1$.
    \item SI Movie 6, ``Continuum\_model\_xi\_5\_0.mp4": same as SI Movie 3 but with $\xi = 5$.
    \item SI Movie 7, ``Continuum\_model\_xi\_infty.mp4": same as SI Movie 3 but with $\xi = +\infty$ (no-slip wall).
    
\end{itemize}
\bibliography{Biblio}